\documentclass[10pt,notitlepage,pra,twocolumn]{revtex4-1}
\usepackage{amsfonts}
\usepackage{amsmath}
\usepackage{amssymb}
\usepackage{graphicx}
\usepackage{float}
\usepackage[toc,page,header]{appendix}
\usepackage{color}

\begin{document}

\title{Efficiency limitation for realizing an atom-molecule adiabatic transfer based on a chainwise system}
\author{Jingjing Zhai, Lu Zhang, Keye Zhang, Jing Qian$^\dagger$ and Weiping Zhang}
\affiliation{Quantum Institute for Light and Atoms, Department of Physics, East China
Normal University, Shanghai 200062, People's Republic of China}

\begin{abstract}
In a recent work we have developed a robust chainwise atom-molecule adiabatic passage scheme to produce ultracold ground-state molecules via photo-associating free atoms [J. Qian {\it et.al.} Phys. Rev. A 81 013632 (2010)]. With the help of intermediate auxiliary levels, the pump laser intensity requested in the atomic photo-association process can be greatly reduced. In the present work, we extend the scheme to a more generalized (2$n$+1)-level system and investigate the efficiency limitation for it. As the increase of intermediate levels and auxiliary lasers, the atom-molecule adiabatic passage would be gradually closed, leading to a poor transfer efficiency. For the purpose of enhancing the efficiency, we present various optimization approaches to the laser parameters, involving order number $n$, relative strength ratio and absolute strength. We show there can remain a limit on the population transfer efficiency given by a three-level $\Lambda$ system. In addition, we illustrate the importance of selecting an appropriate number of intermediate levels for maintaining a highly efficient transfer under mild experimental conditions. 
 
\end{abstract}

\maketitle
\preprint{}

\section{Introduction}
The quest for coherent atom-molecule transfer towards the goal of ultracold molecules has attracted considerable attentions over the last decade \cite{carr09}.
During the conversion process, the stimulated Raman adiabatic passage (STIRAP) treated as a robust technique could strongly correlate the initial and target states by two sequential partially overlapping pulses. An intermediate excited state acquires a negligible population over the whole sequence \cite{bergmann98,Jing07}. The basic concept of STIRAP deals with a three-level $\Lambda$ system in which a dark state is formed by the superposition of two stable states \cite{ling05,ling07,pu07}. 
Due to the quantum interference effect, the middle unstable state is entirely unpopulated in the conversion process.

As for the coherent atom-molecule transfer, using a normal STIRAP based on a $\Lambda$ level structure is experimentally challenging.
An unfavorable Franck-Condon (FC) coefficient to the free-bound photo-association 
leads to the difficulty of finding a proper middle excited state with a good FC overlap with both the delocalized initial atomic state and the tightly-bound molecular state \cite{kerman04,wang04,sage05,salzmann08,viteau08,deiglmayr08}.
Hence, a large number of experimental efforts adopt weakly bound Feshbach molecules instead of free atoms initially, which are then coherently transformed into a tightly-bound state by STIRAP pulses \cite{winkler07,lang08,danzl08,ni08,ospelkaus08,takekoshi14,molony14,park15}. The magnetic Feshbach approach requires suitable levels of atomic species to meet the magnetic resonance \cite{kohler06,chin10,zuchowski10,patel14,wang15}, then it is invalid for a dimer containing one non-magnetic atom and one alkali-metal atom. 
The pioneer work by Aikawa {\it et.al.} first realized a direct conversion of K and Rb atoms into ultracold KRb molecules by photo-association before STIRAP  \cite{aikawa10}, which opens possibilities to coherently manipulate a wide variety of molecules with full-optical methods. Later, S. Stellmer {\it et.al.} prepared atom pairs in a Mott insulator in an optical lattice and attained a largely improved transfer efficiency \cite{stellmer12}.

On the other hand, a multilevel structure is treated as another alternative approach to realize the atom-molecule transfer, due to its high selectivity and controllability for the use of intermediate auxiliary levels \cite{shore91,shore95,malinovsky97,vitanov98a,vitanov98b,vitanov99,sola99,jin04,moller07,shapiro07,qianNJP,qianJPB,niu11,wang12}. 
The first extension to a complex multilevel STIRAP case was proposed by B. W. Shore in 1991 \cite{shore91}, and then more and more attentions are paid to various multilevel schemes, such as the alternating STIRAP in which all the even transitions occur before
the odd ones \cite{shore91}, the straddling STIRAP in which a set of intense pulses corresponding to all intermediate states spans both the Stokes and pump pulses \cite{malinovsky97,sola99}, and the $ (m + n)$ STIRAP which contains $m$-photon pump transitions and $n$-photon Stokes transitions \cite{niu11}. 
Recently Kuznetsova {\it et. al.} suggested a chainwise STIRAP scheme starting from Feshbach molecules. In the scheme, one or two pairs of middle lasers are included for the formation of a robust STIRAP, with minimal influence of intermediate-state decay on the transfer process \cite{kuznetsova08}. 
More recently, we developed a new chainwise STIRAP from photo-associating free atoms \cite{qian10}. In particular, we found appropriate adjustment of the middle laser intensity could equally improve the free-bound FC coefficient, which enables the use of a low pump power for an efficient molecular yield. 
In addition, a four-photon chainwise STIRAP for producing ground-state Cs$_2$ molecules was realized experimentally with Feshbach molecules \cite{danzl10}.

Early works mostly focus on the enhancement of atom-molecule conversion efficiency in this process, e.g. \cite{feng09}. 
Based on our results in ref. \cite{qian10}, intuitively, it shows that if more intermediate lasers are involved, then the intensity of the pump laser as well as the transfer efficiency could be further optimized.
In the present work, we consider the efficiency limit and weakness to a generalized chainwise scheme (as shown in figure \ref{model}). 
Starting from free atoms (like in refs. \cite{javanainen99,mackie00,wynar02,naidon03,kallush08}), we describe the presence of a quasi-dark state and study the adiabatic transfer based on it.
As more and more intermediate energy levels are involved, we show a clear decay for the final molecular probability due to the break of the quasi-dark state. The decay rate   with respect to the order number of system is investigated in detail under the effect of middle-laser parameters.
In general, compared to a three-level $\Lambda$ system, the multilevel chainwise  STIRAP can reduce the intensity of the pump laser at the expense of impaired transfer efficiency, giving rise to a poor molecular production.

\section{Chainwise model and Adiabatic passage} \label{QUASI}
\subsection{Preparing  a quasi-dark state}

\begin{figure}[ptb]
\centering
\includegraphics[width=3.3in,height=2.8in]{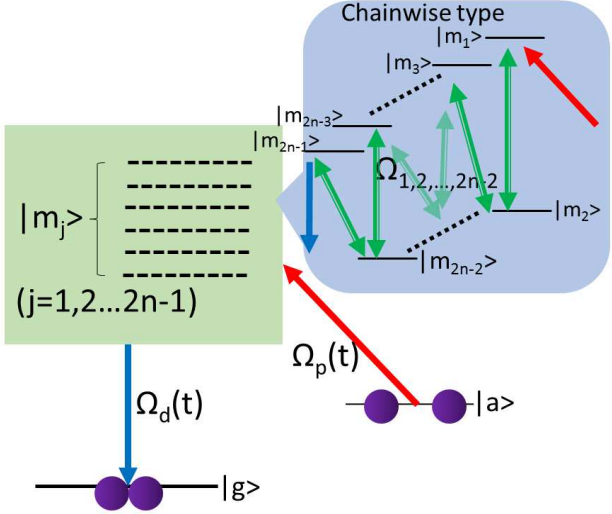}
\caption{(Color online) Schematic of a generalized chainwise system for realizing an adiabatic atom-molecule transfer. $\left\vert a \right\rangle$, $\left\vert m_j \right\rangle$ (j=1,2,...,2n-1) and $\left\vert g \right\rangle$, respectively, represent the initial atomic scattering state, the intermediate molecular states and the target molecular ground state. The initial and the final stage transitions between $\left\vert a \right\rangle$ and $\left\vert m_{1} \right\rangle$, and between $\left\vert m_{2n-1} \right\rangle$ and $\left\vert g
\right\rangle $ are played by 
$\Omega _{p}\left( t\right) $ and $\Omega _{d}\left( t\right) $, respectively, which constitute the pair of STIRAP pulses.
 {\it Insets}: A detailed description of the intermediate levels $\left\vert m_{j} \right\rangle$ coupling with a chain of continuous lasers $\Omega_{1,2,...,2n-2}$ is presented.}
\label{model}
\end{figure}

As presented in figure \ref{model}, the chainwise model we consider contains 2$n$+1 (odd number) energy levels where the initial and the target states are played by the free atomic state $\left\vert a \right\rangle$ and the stable molecular ground state $\left\vert g \right\rangle$. $n$ is treated as the order number of system. Inset presents a detailed description for the intermediate energy levels $\left\vert m_{j} \right\rangle$(j=1,2,...2n-1) coupling with the middle lasers $\Omega_{1,2,...,2n-2}$. There are $n$ middle excited states and $n-1$ middle metastable states. 
Transition between $\left\vert a \right\rangle$ and $\left\vert m_1 \right\rangle$ is performed by pulse $\Omega_p(t)$, between $\left\vert m_{2n-1} \right\rangle$ and $\left\vert g \right\rangle$ by pulse $\Omega_d(t)$.  Two neighboring states $\left\vert m_{j} \right\rangle$ and $\left\vert m_{j+1} \right\rangle$ are coupled by a time-independent laser $\Omega_j$.

Here we consider a spatially uniform condensate described by the annihilation operators $\hat{\psi_{i}}$ and creation operators $\hat{\psi_{i}^{\dagger}}$ ($i=a$ for atoms and $i=m_{j},g$ for molecules), and the average particle density is $n_{T}$. The Hamiltonian of the system is given in a generalized form \cite{heinzen00},
\begin{equation}
\frac{\hat{H}}{\hbar} =(\frac{\Omega _{p}^{\prime}(t)}{2}\hat{\psi}_{m_{1}}^{\dagger }%
\hat{\psi}_{a}^{2}+\frac{\Omega _{d}(t)}{2}\hat{\psi}_{g}^{\dagger }\hat{\psi}%
_{m_{2n-1}}+\sum_{j=2}^{2n-1}\frac{\Omega _{j-1}}{2}\hat{\psi}%
_{m_{j}}^{\dagger }\hat{\psi}_{m_{j-1}})+H.c.  \label{Ham}
\end{equation}%
where $\Omega_p^{\prime}(t)=\Omega_p(t)/\sqrt{n_T}$ accounts for two atoms associating into one dimer molecule. The first, second and third terms in the right side of Eq. (\ref{Ham}) correspond to the interplay of atoms and molecules in $\left \vert m_1\right\rangle$, ground-state molecules and molecules in $\left \vert m_{2n-1}\right\rangle$, two middle-state molecules $\left \vert m_{j}\right\rangle$ and $\left \vert m_{j-1}\right\rangle$, respectively. Here, we ignore the two-body $s$-wave collisions without loss of our main physics \cite{pu07}. 
The detunings are assumed to be zero by the assumption of resonant laser excitations \cite{shore91}, however, if the system contains even number of states, a modest middle state detuning is required for realizing an adiabatic evolution \cite{oreg92}. Besides, the effect of the laser noise from the fluctuations of laser-field amplitude and phase on the transfer efficiency will be discussed in section \ref{realization}.

Hamiltonian (\ref{Ham}) has the nonlinear term (the first term) so that it can not be diagonalized directly. We introduce the Heisenberg motional equation
\begin{equation}
	 i\hbar \partial _{t}{\mathbf{\hat{\Psi}}}\left( t\right) =[\hat{H}-i\hbar\sum_{j=1}^{2n-1}\gamma_{j}\hat{\psi}%
_{m_{j}}^{\dagger }\hat{\psi}_{m_{j}},{\mathbf{\hat{\Psi}}}\left( t\right) ]
\label{motion}
\end{equation}
to describe the dynamics of the system, in which the field operator is given by 
\begin{equation}
\hat{\mathbf{\Psi}}(t)=(\hat{\psi}
_{a}(t),\hat{\psi}_{m_{1}}(t),\cdots ,\hat{\psi}_{m_{2n-1}}(t),\hat{\psi}
_{g}(t))^{T}
\end{equation}
and the spontaneous decay rate $\gamma_j$ of the middle states $\left \vert \hat{\psi}_{m_j} \right \rangle$ are introduced phenomenologically.  When a pair of counter-intuitive laser pulses $\Omega_p(t)$ and $\Omega_d(t)$ is performed, the system will adiabatically evolute into a steady state under the effect of laser driving and spontaneous dissipation. 
This steady state is treated as a dark eigenstate.
 Note that in the present scheme, we only find a quasi-dark eigenstate. The name "quasi-dark" originates from non-zero population on middle metastable states $\left\vert m_{2,4,\cdots,2n-2}\right\rangle$.

The steady state population can be solved as follows.
Firstly, let $\partial _{t}\mathbf{\hat{\Psi}}\left( t\right) =0$ and replace the field operators $\hat{\psi}_i(t)$ ($\hat{\psi}_{i}^{\dagger}(t)$) by its $c$ number values $\sqrt{n_T} \phi_{i}(t)$ ($\sqrt{n_T} \phi_{i}^{*}(t)$) under the mean-field treatment. Secondly, by ignoring the decays and considering the conservation of the total particle number 
\begin{equation}
|\phi_a|^2+2|\phi_g|^2+2\sum_{j=1}^{2n-1}|\phi_{m_j}|^2=1,
\end{equation}
we can solve for the steady-state population. The amplitude $\phi_i$ in each bare state of the quasi-dark eigenstate can be solved as
\begin{eqnarray}
\phi _{a}& =&\sqrt{\frac{2}{1+\sqrt{1+8(\eta _{1}\chi )^{2}}}} \label{cwCPTa}\\
\phi _{g}& =&\frac{(-1)^{n}2\eta _{1}\chi }{1+\sqrt{1+8(\eta _{1}\chi )^{2}}}\label{cwCPTb}\\
\phi _{m_{2}}& =&-\xi \phi _{a}^{2}, \label{cwCPTc}\\
\phi _{m_{2j-2}}& =&(-1)^{j-1}\xi \frac{\Omega _{2}\cdots \Omega _{2j-4}}{%
\Omega _{3}\cdots \Omega _{2j-3}}\phi _{a}^{2},\qquad\mbox{for}\qquad
j \geq 3 \label{cwCPTd}\\
\phi _{m_{1,3,...,2n-1}}&=&0
\end{eqnarray}%
where $\xi(t)=\Omega _{p}(t)/\Omega_{1}$ and $\chi(t)=\Omega _{p}(t)/\Omega _{d}(t)$ represent the relative strength of lasers and 
\begin{equation}
	\eta_1=\prod_{j=1}^{n-1}\alpha_j,\alpha _{j}=\Omega _{2j}/\Omega _{2j-1}
\label{ratio}
\end{equation}
presents a new important parameter involving the effect of all middle-state lasers. 
If $\eta_1=1$ and $n=1$, our solutions (\ref{cwCPTa}) and (\ref{cwCPTb}) could reduce into the exact dark state form in a three-level $\Lambda$ system, e.g., see \cite{pu07}. 
In (\ref{cwCPTa}) and (\ref{cwCPTb}) when the value of $\eta _{1}\chi (t)$ changes
from $0$ to +$\infty $ with $\Omega _{p}(t)$ and $\Omega _{d}(t)$ properly prepared
in a counter-intuitive order (note that $\eta_1$ is a constant), a complete atom-molecule transfer will take place between states $\left \vert a \right \rangle$ and $\left \vert g \right \rangle$, as shown in figure \ref{adia}(a). 
Compared with a three-level $\Lambda$ case, $\eta_1$ acting as a new control knob can deeply affect the atom-molecule conversion rate in the quasi-dark state case here. 
As illustrated by figure \ref{adia}(a), if $\eta_1<$1 (red dash curves) the atoms convert into molecules at a later time compared to the case of $\eta_1=1$ (blue solid curves), and if $\eta_1>$1 (black dash-dotted curves) their conversion occurs at an earlier time. Especially if $\eta_1$=100 (green dotted curves) we observe the conversion occurs at a very early time around $t=500\mu s$.
The result from the three-level case (blue curves with circles) perfectly coincides with  the one from the $\eta_1$=1 case.

In addition, the non-zero population probabilities on the intermediate metastable states $\left \vert m_{2}\right\rangle$ and $\left \vert m_{2j-2}\right\rangle$ [see (\ref{cwCPTc}) and (\ref{cwCPTd})]
could lead to a severe loss due to the spontaneous decays. However, it is fortunate to find that by using a very small $\xi$, i.e. letting $\Omega_1\gg\Omega_p$, the molecular occupations on these states can gain a large suppression. Ensuring $\Omega_1\gg\Omega_p$ is therefore significantly important for achieving an efficient transfer in chainwise schemes. In our calculations, $\Omega_1=10\Omega_p^{max}$ is used.

\subsection{Adiabatic condition of the quasi-dark state} \label{adiacond}

Generally speaking, the presence of a quasi-dark eigenstate can not ensure an exact adiabatic passage, since any small perturbations from the surroundings may break its adiabaticity \cite{meng14}. To qualify the adiabaticity, we adopt a typical way by adding small perturbation components $\delta\psi_i(t)$ to the time-dependent quasi-dark state components $\phi_{i}(t)$. If $\delta\psi_i(t)$ converge with time the state is stable, otherwise, it is unstable.
We linearize the Heisenberg motional equation (\ref{motion}) by assuming $\psi_{i}(t)=\phi_{i}(t)+\delta \psi_{i}(t)$. After some arrangements, we can arrive at a set of coupled equations:
\begin{equation}
i\partial_{t}\delta{\mathbf{\Psi}}\left( t\right) =({\mathbf{M}}\left( t\right)
-i\mathbf{\Gamma})\delta\mathbf{\Psi}\left( t\right) -i\partial_{t}\mathbf{%
\Phi}\left( t\right)  \label{pert}
\end{equation}
where the matrice $\delta\mathbf{\Psi}\left( t\right) $ and $\partial_{t}\mathbf{%
\Phi}\left( t\right) $ take forms of 
\begin{eqnarray}
\delta\mathbf{\Psi}(t)&=&(\delta\psi_{a},\delta\psi_{m_{1}},\delta\psi_{m_{2}},%
\cdots,\delta\psi_{m_{2n-1}},\delta\psi_{g})^{T} \label{elema}\\
\partial_{t}\mathbf{\Phi}\left( t\right) &=&(\dot{\phi}_{a},0,\gamma_{2}%
\phi_{m_{2}}+\dot{\phi}_{m_{2}},\cdots,\gamma_{2n-2}\phi_{m_{2n-2}}+\dot{%
\phi }_{m_{2n-2}},0,\dot{\phi}_{g})^{T} \label{elemb}
\end{eqnarray}
and $\mathbf{M}$ that the matrix of $(2n+1)\times(2n+1)$ dimensions represents the system-light coupling strengths, whose non-zero elements are merely given by
$\mathbf{M_{1,2}}=\mathbf{M_{2,1}}=\Omega_{p}\phi_{a},$ $\mathbf{M_{2n,2n+1}}%
=\mathbf{M_{2n+1,2n}}=\Omega_{d}/2$, and 
$\mathbf{M_{j,j+1}}=\mathbf{M_{j+1,j}}%
=\Omega_{j-1}/2 $. 
The matrix $\mathbf{\Gamma}$ in (\ref{pert}) represents the spontaneous decay rate from the intermediate states (the initial and target states are long-lifed). It is diagonal with nonzero elements
$\mathbf{\Gamma_{jj}}=\gamma_{j-1}$($j=2,3,\cdots2n-1$). 
Diagonalizing $\mathbf{M}$ leads to a complete set of eigenenergy and eigenvector. The eigenenergy of the quasi-dark state $\left \vert \omega_0 \right \rangle$ is zero, i.e., $\omega_0=0$. Besides, we focus on two nearest neighboring eigenstates with respect to $\left \vert \omega_0 \right \rangle$, whose eigenenergies are
\begin{equation}
\omega_{\pm}=\pm\frac{\Omega_{eff}}{2\beta_{n-1}},
\label{omegapm}
\end{equation}
and the eigenvectors are given by
\begin{equation}
\left\vert \omega_{\pm}\right\rangle =\frac{1}{\sqrt{2}}(\frac{2\Omega
_{p}\phi_{a}\eta_{1}}{\Omega_{eff}},\pm\frac{\eta_{1}}{\beta_{n-1}}%
,0,\mp\frac{\eta_{2}}{\beta_{n-1}},\cdots,0,\pm\frac{1}{\beta_{n-1}},\frac{\Omega_{d}}{\Omega_{eff}})^{T}
\label{eigen_chain}%
\end{equation}
where the effective Rabi frequency is defined as $\Omega_{eff}=\Omega_{d}^{1/2}(\Omega_{d}^{2}+8\eta_{1}^{2}%
\Omega_{p}^{2})^{1/4}$. For a three-level $\Lambda$ system, 
$\Omega_{eff}^{\Lambda}=\Omega_{d}^{1/2}(\Omega_{d}^{2}+8\Omega_{p}%
^{2})^{1/4}$. In Eqs. (\ref{omegapm}) and (\ref{eigen_chain}), we define $\beta_{n-1}$ as another control knob merely for the chainwise cases. $\beta_{n-1}=\sqrt{\sum_{k=1}^{n-1}\eta_k^2+1}$ which is related to all $\eta_k$ with $\eta_{k}={\prod}_{j=k}^{n-1}\alpha_{j}(k=1,2,\cdots,n-1)$ from stage $k$.

Comparing to the case of a three-level $\Lambda$ system, in which eigenenergies and eigenvectors 
\begin{equation}
\omega_{\pm}^{\Lambda}=\pm\frac{\Omega_{eff}^{\Lambda}}{2},\left\vert \omega_{\pm}^{\Lambda}\right\rangle =\frac{1}{\sqrt{2}}\left(
\frac{2\Omega_{p}\phi_{a}}{\Omega_{eff}^{\Lambda}},\pm1,\frac
{\Omega_{d}}{\Omega_{eff}^{\Lambda}}\right)^{T}
\label{eigen_l}%
\end{equation}
are obtained, 
we find the presence of $\beta_{n-1}$ ($\beta_{n-1}$ $>1$ from its definition) in (\ref{omegapm})
would lead to $|\omega_{\pm}|$ $<$ $|\omega_{\pm}^{\Lambda}|$
even if $\eta_1=1$ is used. That fact means, in the chainwise picture, there are two non-adiabatic eigenstates $\left \vert \omega_{\pm} \right \rangle$ locating closer to the quasi-dark state $\left \vert \omega_{0} \right \rangle$. So that the adiabatic passage based on a chainwise type system is easier to break
due to its smaller energy separation between $\left\vert \omega_{0}\right\rangle$ and $\left\vert \omega_{\pm}\right\rangle$ (the separation of $\left\vert \omega_{0}^{\Lambda}\right\rangle$ and $\left\vert \omega_{\pm}^{\Lambda}\right\rangle$ is relative large).

\begin{figure}[ptb]
\centering
\includegraphics[width=2.78in,height=3.1in]{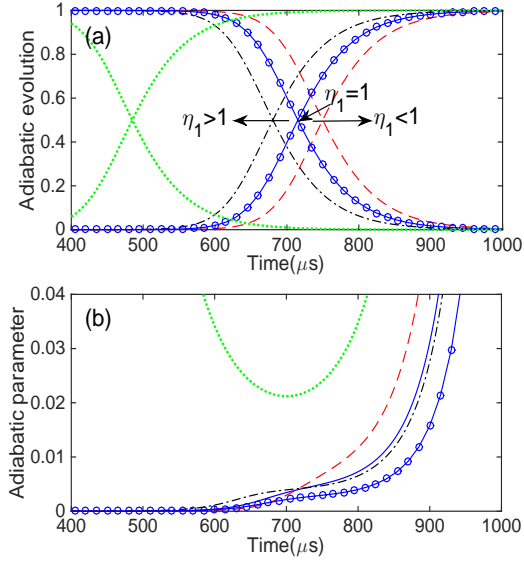}
\caption{(Color online) (a) The atom-molecule adiabatic evolution $|\phi_a(t)|^2$ (1 $\rightarrow$ 0) and $2|\phi_g(t)|^2$ (0 $\rightarrow$ 1) in the quasi-dark eigenstate. We choose $\eta_1$=100 (green-dotted curve), 2.0 (black dash-dotted curves), 1.0 (blue solid curve), 0.5 (red dashed curve).  (b) The corresponding time-dependent adiabatic parameter $r(t)$.
The blue curve with circles presents the results from a three-level $\Lambda$ system. The order number $n=2$ is considered. Other parameters are defined in the text.}
\label{adia}
\end{figure}

To further quantify the influence of middle chainwise levels, one resorts to calculate the population on two non-adiabatic eigenstates $\left\vert \omega_{\pm}\right\rangle$, because they are the most probable states to accumulate except $\left\vert \omega_{0}\right\rangle$. We consider 
$\delta \mathbf{\Psi}(t)={\sum}_{u=\pm}
c_{u}(t)\left\vert \omega_{u}^{C}(t)\right\rangle $ where $c_{u}$ 
is the state projection. 
Other higher-order excited eigenstates are safely ignored, accounting for their eigenenergies $\xi^{-1}$ ($\xi \ll 1$) larger than $|\omega_{\pm}|$, as well as weak couplings to $\left \vert\omega_{\pm} \right \rangle$ and $\left \vert \omega_{0} \right \rangle$.
A detailed description for ignoring these high-order eigenstates has been presented elsewhere \cite{ling07}.
Finally, the adiabatic parameter can be given by
\begin{equation}
r(t)=\frac{1}{2}\sqrt{\left\vert c_{+}\right\vert ^{2}+\left\vert
c_{-}\right\vert ^{2}}
\end{equation}
describing the non-adiabatic excitation probabilities in two non-dark eigenstates $\left\vert \omega_{\pm}\right\rangle$. Obtaining $r(t)$ is analytically derived as
\begin{equation}
r(t)  =\frac{4\eta_{1}|\dot{\chi}|\beta_{n-1}^{2}}{\Omega_{d}^{2}(1+\sqrt
{1+8\eta_{1}^{2}\chi^{2}})(1+8\eta_{1}^{2}\chi^{2})^{3/4}}\sqrt{\gamma_{1}%
^{2}+(\omega_{+})^{2}} 
\label{keyb}%
\end{equation}
For comparison we also present the adiabatic parameter $r^{\Lambda}(t)$ of a three-level $\Lambda$ system, which is
\begin{equation}
r^{\Lambda}(t) =\frac{4|\dot{\chi}|}{\Omega_{d}^{2}(1+\sqrt{1+8\chi^{2}%
})(1+8\chi^{2})^{3/4}}\sqrt{\gamma_{1}^{2}+(\omega_{+}^{\Lambda})^{2}%
}. \label{keya}
\end{equation}

In arriving at (\ref{keyb}) and (\ref{keya}), we assume a same spontaneous decay rate for all intermediate excited states that is $\gamma_{1,3\cdots2n-1}=\gamma_{1}$.
Equation (\ref{keyb}) is a key result in the chainwise transfer process which gives important information about the adiabatic condition of this complex quasi-dark state. 
Obtaining a larger $r(t)$ value means the chainwise system is easier to be non-adiabatically excited off the quasi-dark state $\left\vert\omega_0\right\rangle$, although it is initially prepared in $\left\vert\omega_0\right\rangle$. 
Obviously, it is worth to stress that the condition $r(t) > r^{\Lambda}(t)$ is always met because $r(t)$ is proportional to $\beta_{n-1}$ and $\beta_{n-1} > 1$ according to its definition.

A numerical identification is presented in figure \ref{adia}(b) where we have shown $r(t)$ at $\eta_1$=100 (green dotted curve), 2.0 (black dash-dotted curve), 1.0 (blue solid curve), 0.5 (red dashed curve) and $r^{\Lambda}(t)$ (blue curve with circles), respectively. We observe that by increasing $\eta_1$ from 0.5 to 2.0, $r(t)$ becomes closer to $r^{\Lambda}(t)$, but $r(t) > r^{\Lambda}(t)$ is always satisfied. If $\eta_1$ is significantly large, coming to 100 (green dotted curve), $r(t)$ shows a great change and increases by orders of magnitude. 
That displays the importance of an appropriate $\eta_1$ while preparing middle lasers. In addition, we also realize another key parameter $\beta_{n-1}$ in (\ref{keyb}) that strongly relies on the choice of order number $n$. By combining $\eta_1$ and $\beta_{n-1}$, we could optimize the adiabatic parameter $r(t)$ for achieving a robust atom-molecule transfer based on chainwise systems. Meanwhile, it allows us to understand the efficiency limit and the weakness of the chainwise type schemes.

\section{Optimization of adiabatic parameter}

To ensure a robust atom-molecule transfer in a multilevel chainwise system, we have proposed an idea to adjust the relative strength $\alpha_{j}$ of two middle neighbor lasers, in order to compensate the free-bound FC coefficient \cite{qian10}. Presently, we aim to extend our former results to a more generalized chainwise system and study its efficiency limit as more middle energy levels are included. In section \ref{QUASI}, we have developed a quasi-dark eigenstate approach and solved an adiabatic parameter $r(t)$ to quantify the performance of the adiabatic passage \cite{jia12}. If $r(t)$ is small, the atom-molecule conversion can adiabatically follow the quasi-dark eigenstate and a high transfer efficiency will be achieved.
We mainly focus on how to minimize $r(t)$ by
two controllable quantities $\eta_1$ and $\beta_{n-1}$ below. From the definition of $\eta_1$ and $\beta_{n-1}$ we see they two can be tuned by $\alpha_{j}$ and $n$. $\alpha_j$ is the relative strength of two middle lasers and $n$ is the order number of system. Both of them are experimentally adjustable. In the calculations, we assume the pump laser $\Omega_p(t)$ and the Stokes laser $\Omega_d(t)$ are Gaussian-shaped, which are
\begin{eqnarray}
\Omega_p(t) &=& \Omega_p^{max}exp(-(t-t_{p})^2/T^2), \\
\Omega_d(t) &=& \Omega_d^{max}exp(-(t-t_{d})^2/T^2),
\end{eqnarray}
with the peak intensities $\Omega_p^{max}=2\times10^6$Hz, $\Omega_d^{max}=2\times10^7$Hz, the central positions $t_p=800\mu$s, $t_d=400\mu$s,  and the pulse width $T=200\mu$s. The spontaneous emissions are $\gamma_{1,3\cdots2n-1}=10^6$Hz, $\gamma_{2,4\cdots2n-2}=10^4$Hz.

\subsection{ A constant $\alpha_0$ with respect to $n$} \label{secord}
As seen from equation (\ref{ratio}), we have defined the ratio $\alpha_j$ for the relative strength between neighboring even-order and odd-order intermediate lasers. For simplicity, we first consider a same ratio i.e. $\alpha_j$=$\alpha_0$. We start from (\ref{keyb}) and rewrite it by replacing $\alpha_j$ by $\alpha_0$. Notice that in the system-light interaction where $\gamma_1\gg |\omega_{\pm}|$ is met under the given parameters, then $r(t)$ can reduce into
\begin{equation}
r(t)\propto\frac{\gamma_1\beta_{n-1}^2\left\vert \dot{\chi(t)} \right\vert/\Omega_{d}^2(t)}{2^{\frac{7}{4}}\alpha_0^{\frac{3}{2}(n-1)}\chi^{\frac{5}{2}}}
\label{rappro}
\end{equation}
where $\beta_{n-1} =\sqrt{\left( 1-\alpha _{0}^{2n}\right) /\left( 1-\alpha
_{0}^2\right) }$ for $\alpha _{0}\neq 1$. By taking the derivative of (\ref{rappro}) with respect to $\alpha_{0}$, it is easy to obtain an optimal $\alpha_{0}$ with which $r(t)$ attains its local minimum point. Thus, $\alpha_0$ can be solved from an equality

\begin{equation}
\alpha _{0}^{2n}=\frac{3\left( n-1\right) -\alpha _{0}^{2}\left( 3n+1\right) 
}{\alpha _{0}^{2}\left( n-1\right) -\left( n+3\right) }.  
\label{cond}
\end{equation}
 
\begin{figure}[ptb]
\centering
\includegraphics[width=2.72in,height=1.89in]{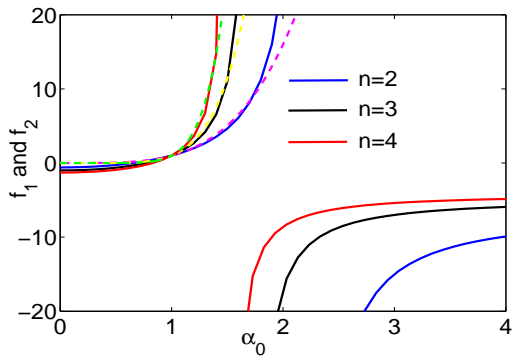}
\caption{(Color online) Functions $f_{1}(\alpha_{0})$ and $f_{2}(\alpha_{0})$ versus $\alpha_0$ with respect to the order numbers $n$. For $n$=2, 3 and 4, $f_{1}$ is marked by the magenta dashed curve, yellow dashed curve and green dashed curve, respectively; and $f_{2}$ by blue solid curve, black solid curve and red solid curve, respectively. Optimal $\alpha_0$ values are obtained by the crossing of $f_1$ and $f_2$ on the positive part of $f_2$, which are accordingly given by $\alpha_0^{(n=2)}$=1.73, $\alpha_0^{(n=3)}$=1.51, $\alpha_0^{(n=4)}$=1.41.}
\label{levelnumberorder}
\end{figure}

To solve (\ref{cond}) we define two functions, $f_{1}(\alpha_0)=\alpha _{0}^{2n}$ and $f_{2}(\alpha_0)=\frac{3\left( n-1\right) -\alpha _{0}^{2}\left( 3n+1\right) 
}{\alpha _{0}^{2}\left( n-1\right) -\left( n+3\right) }  $, and plot them in figure \ref{levelnumberorder}.
From left to right we choose $n$=4 (green dashed and red curves), 3 (yellow dashed and black curves) and 2 (magenta dashed and blue curves). For all $n$ values, $\alpha_0$=1 is a common root where $f_1(1)=f_2(1)$. Besides, for a decided $n$, we find there always exists another real root on the positive part of $f_2$, which satisfies $f_1(\alpha_0^{(n)})=f_2(\alpha_0^{(n)})$ and $\alpha_{0}^{(n)}\in (1,2)$.
With the increase of $n$, $\alpha_0^{(n)}$ becomes small and close to $\alpha_0$=1.

\begin{figure}[ptb]
\centering
\includegraphics[width=3.12in,height=2.19in]{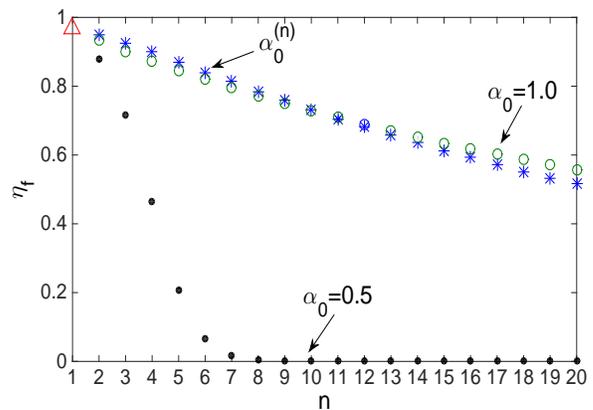}
\caption{(Color online) The final atom-molecule conversion efficiency $\eta_f$ versus the order numbers $n$ by using $\alpha_0$=$\alpha_0^{(n)}$(blue stars), $\alpha_0$=1.0 (green circles) and $\alpha_0$=0.5 (black dots), respectively. $\eta_f^{\Lambda}$ is given by a red triangle at $n=1$, describing the transfer efficiency in a three-level system. In the calculation, the Rabi frequencies of the middle odd-number lasers are $\Omega_{2j-1}=2\times10^7$Hz. }
\label{neffi}
\end{figure}

To verify that using $\alpha_0^{(n)}$ could increase the transfer efficiency, we let $\eta_f=2|\psi_g(t\to \infty)|^2$, describing the final molecular probability, and study its relationship to $n$. We choose $\alpha_0=$$\alpha_0^{(n)}$ (optimal), 1.0 and 0.5 in the calculation. For comparison we also plot $\eta_f^{\Lambda}$ (marked by a red triangle at $n=1$) from the three-level $\Lambda$-type system by using the same pulses $\Omega_p(t)$ and $\Omega_d(t)$. 
The results are presented in Figure \ref{neffi}. 
Generally speaking, as $n$ increases, $\eta_f$ always decreases with different decay rates depending on $\alpha_0$. In particular, if $\alpha_0$=0.5 which is a non-optimal $\alpha_0$ value, $\eta_f$ rapidly falls into zero after $n > 7$, however, with the use of optimal $\alpha_0^{(n)}$ or $\alpha_0=1$, we see the decay rate of $\eta_f$ with respect to $n$ becomes much slowly. Especially in the case of $n=20$ the final molecular probability remains as high as 0.6. 
That means there would be a great enhancement for the yield of molecules in chainwise systems by choosing an optimal $\alpha_0$.

In addition, we also find the transfer efficiency based on a three-level $\Lambda$ system is maximal, i.e. $\eta_f^{\Lambda}$=0.967$>\eta_f$, denoted by a red triangle at $n=1$ in figure \ref{neffi}. That result is reasonable because auxiliary middle levels in chainwise schemes would strongly affect the coherence between the initial and target stable states, while in a three-level $\Lambda$ system an exact dark eigenstate directly correlates the two stable states $\left \vert a\right\rangle$ and $\left \vert g\right\rangle$ which is easy to keep coherent and adiabatic. Thereby, no matter how to adjust $\alpha_0$ between middle lasers it is still impossible to overcome the transfer efficiency limit $\eta_f^{\Lambda}$ given by the three-level system. This finding has a perfect agreement with the adiabaticity discussion in subsection 2\ref{adiacond}.

\subsection{A varying $\alpha_j$ when n=3} \label{relative}

In this subsection we choose the case $n=3$ as an example to see whether a variable $\alpha_j$ could further make the efficiency $\eta_f$ a step closer to the limit $\eta_f^{\Lambda}$. 

\begin{figure}[ptb]
\centering
\includegraphics[width=3.02in,height=1.99in]{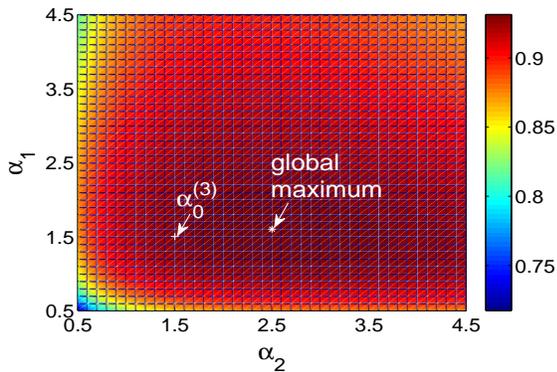}
\caption{(Color online) The final atom-molecule transfer efficiency $\eta_f$ versus ($\alpha_1$, $\alpha_2$) when $n=3$ is considered. $\alpha_0^{(3)}=(1.51,1.51)$ is pointed by a cross, the global maximum efficiency $\eta_f^{(max)}$ is pointed by a star with $(\alpha_1,\alpha_2)=(1.6,2.5)$.}
\label{popu2}
\end{figure}

Note that there exists two variables $\alpha_1$ and $\alpha_2$ when $n=3$. 
In figure \ref{popu2} we present $\eta_f$ in the parameter space ($\alpha_1, \alpha_2$).
When $\alpha_1=\alpha_2=1.51$, as solved from Eq. (\ref{cond}), we obtain $\eta_f=0.926$. When $\alpha_1$ and $\alpha_2$ are both adjustable as plotted in figure \ref{popu2}, we find the global maximum located at $(\alpha_1,\alpha_2)=(1.6,2.5)$ rather than $(\alpha_1,\alpha_2)=(1.51,1.51)$.
 The maximal transfer efficiency solved is $\eta_f^{(max)}=0.931$, which is only slightly larger than 0.926 by using $\alpha_0^{(3)}=1.51$; but, it is still smaller than $\eta_f$=0.95 solved from the case of $n=2$. Thereby, with a fully tunable $\alpha_j$, a gradual decaying of $\eta_f$ remains with the increase of middle energy levels. For a determined $n$, i.e. the number of middle energy levels is fixed, the atom-molecule transfer efficiency would be improved by an appropriate control of $\alpha_j$. Nevertheless, while comparing to the $(n-1)$ case, $\eta_f$ is still decreasing.

A brief conclusion for subsections 3\ref{secord} and 3\ref{relative} is represented that in the optimization of $r(t)$, we find within the chainwise multilevel systems, $\eta_f$ would always be decreasing as the increase of middle energy levels. That decreasing rate can be deeply reduced by optimizing the strength ratio $\alpha_j$ between the two middle neighboring lasers. We also find it is impossible to overcome the transfer efficiency limit given by a typical three-level $\Lambda$ system, by the use of changing $\alpha_j$ in middle lasers.

\section{Optimization of odd-number lasers}

\begin{figure}[ptb]
\centering
\includegraphics[width=3.42in,height=2.29in]{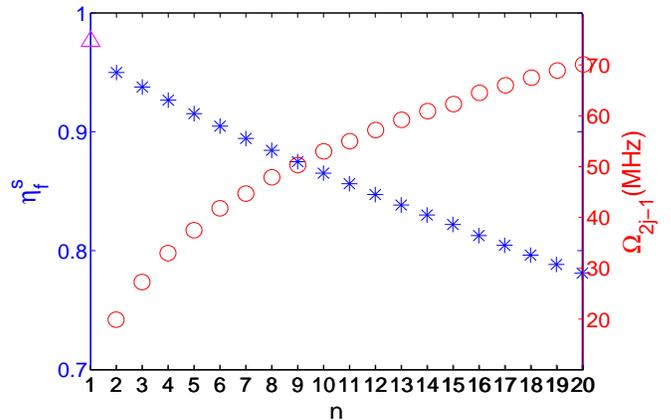}
\caption{(Color online) The saturate atom-molecule conversion efficiency $\eta_f^s$ (marked by blue stars) with order numbers $n$ by using different middle laser strengths $\Omega_{2j-1}$ (odd-number lasers, marked by red circles). The Rabi frequencies of the even-number lasers are $\Omega_{2j}=\alpha_0^{(n)}\times\Omega_{2j-1}$. The red triangle also shows the saturate efficiency based on a three-level $\Lambda$ system.
}
\label{exact}
\end{figure}

Note that in (\ref{keyb}) the adiabatic parameter $r(t)$ only depends on the relative strength $\alpha_j$, rather than the absolute strength $\Omega_j$ of middle lasers. We have studied the effect of $\alpha_j$ on the final molecule population in subsections 3\ref{secord} and 3\ref{relative}. 
On the other hand, according to equations (\ref{cwCPTc})-(\ref{cwCPTd}), it needs the condition $\Omega_{1}\gg\Omega_{p}$ to ensure the adiabaticity and coherence of the quasi-dark state, 
otherwise, the atom-molecule adiabatic passage may be easily broken owing to the spontaneous emission from those intermediate metastable states $\left\vert m_{2,4,\cdots,2n-2} \right\rangle$. 
In this subsection, we pay special attentions to the smallest value of Rabi frequency required for supporting a saturate transfer efficiency. 
In the calculations, we assume all odd-number Rabi frequencies $\Omega_{2j-1}$ are same, and the even-number lasers are $\Omega_{2j}=\alpha_{0}^{(n)}\times\Omega_{2j-1}$.

Figure \ref{exact} shows the relationship between saturate transfer efficiency $\eta_f^s$ and the corresponding $\Omega_{2j-1}$. When $n$ grows, it is obvious to see, the decay rate of $\eta_f^s$ is much smaller than that of $\eta_f$ in figure \ref{neffi}. That is because in the calculations, we also enhance the odd-number Rabi frequency $\Omega_{2j-1}$ to meet for its saturate efficiency $\eta_f^s$. 
For comparing results at a same level, in the plot we choose $\eta_f=\eta_f^s$ with $n=2$ and $\Omega_{2j-1}=2\times10^7$Hz. When $n$ increases, we find the required $\Omega_{2j-1}$ for achieving a saturate efficiency is always growing, from $\Omega_{2j-1}=2\times10^7$Hz at $n=2$ to $\Omega_{2j-1}=7\times10^7$Hz at $n=20$. Accordingly, $\eta_f^s$ approaches as high as 0.80 at $n=20$ with a large middle laser intensity. If $\Omega_{2j-1}=2\times10^7$Hz, the optimal $\eta_f$ is only around 0.60 at $n=20$.
These facts lead to the requirement of choosing relatively intensive middle lasers when more and more middle energy levels are considered. Via the intensive middle lasers, the final transfer efficiency would be deeply enhanced.

\section{Experimental Realization} \label{realization}

In experimental, a multilevel chainwise system can be realized in, e.g., a seven-state bosonic Rb$^{87}$-Rb$^{87}_2$ conversion. In the first step, the atomic scattering state can be coupled to an electronically excited molecular state $\left\vert0_g^-(\sim5^2S_{1/2}+5^2P_{1/2});v,J=0\right\rangle$ \cite{wynar00} via a photon-association laser $\Omega_p$, in which the free-bound FC coefficient can be estimated by $I_{f-b}\sim3\times10^{-14}$m$^{3/2}$ \cite{naidon03} . 
The target molecular ground state $X^1\sum_u^+(v=0,J=0)$ can then be reached by using a chain of laser transitions (see the inset of Fig. \ref{model}) given in Table \rm{I} of ref. \cite{kuznetsova08}. 
In subsequent laser transitions, the bound-bound FC factors are favorable, typically $I_{b-b}\sim 10^{-1}$. In the calculations, we consider the average particle density $n_{T}=10^{20}$m$^{-3}$, the pump laser $\Omega_p^{max}=2\times 10^6$Hz, the Stokes laser $\Omega_d^{max}=2\times 10^7$Hz, and the odd-number lasers $\Omega_{2j-1}=2\times 10^7$Hz. We could estimate the pump laser intensity $I_p=2c\epsilon_0(\hbar\Omega_{p}^{max}/I_{f-b}\sqrt{n_T}u)\sim1000$W/cm$^2$,
$u$ is the transition dipole moment. Correspondingly, other laser intensities are 
 $I_d\sim0.5$W/cm$^2$, and $I_{2j-1}\sim 19$W/cm$^2$. These parameters are all available by current experiment techniques.

In addition, since a chainwise system contains a series of lasers, that the influence of random fluctuations of these laser amplitudes and phases on the atom-molecule transfer should be considered. Because laser fluctuations would dephase the atomic coherences. Stabilizing the lasers is a natural approach for reducing fluctuations, e.g., a compact and robust diode-laser system with high passive stability has been recently realized in experiment \cite{kirilov15}, which can be applied to the atom-molecule transfer \cite{danzl10}. Alternatively, there exists optimum laser Rabi frequencies that can balance the non-adiabaticity and the noise losses in STIRAP procedure \cite{yatsenko14}. Increasing the laser amplitude would enhance the phase fluctuations although the non-adiabaticity can be strongly overcome. Note that in our calculations, the laser Rabi frequencies required are appropriate and optimal.

\section{Conclusions}

We theoretically investigate the efficiency limit of an atom-molecule transfer in a chainwise multilevel adiabatic passage. In the transfer process, a pair of Gaussian-shaped pump and Stokes STIRAP pulses as well as a chain of continuous lasers correlate the initial and target stable states, resulting in the yield of a quasi-dark eigenstate. The adiabaticity in such a state can be characterized by a generalized adiabatic parameter $r(t)$ which enables the study of efficiency optimization. 
We perform a detailed description for optimizing the adiabatic parameter and pay special attentions to the order number and laser Rabi frequencies, which are fully controllable parameters in experiments. Compared to a three-level $\Lambda$ case in which the efficiency can be unitary in principle, the final transfer efficiency is found to decrease when more and more middle energy levels are involved.

For realizing a robust atom-molecule transfer in a chainwise system, it is important to consider a proper number of middle levels and proper laser Rabi frequencies. Then, not only the selectivity of energy levels is greatly improved, but also the coherence and high transfer efficiency can be maintained.
Finally, we study the experimental feasibility for such a multilevel chainwise adiabatic passage by considering a realistic seven-level transition in Rb$^{87}$ atoms. The effect of laser fluctuations on STIRAP is also briefly discussed.

\section{Funding Information}

This work was supported by the NSFC under Grants No. 11474094, No. 11104076, the
Specialized Research Fund for the Doctoral Program of Higher Education No.
20110076120004.


\begin{thebibliography}{99}


\bibitem{carr09} L. D. Carr, D. DeMille, R. V. Krems and J. Ye,"Cold and ultracold molecules: science, technology and applications," New J. Phys. \textbf{11} 055049 (2009).

\bibitem{bergmann98} K. Bergmann, H. Theuer and B. W. Shore,"Coherent population transfer among quantum states of atoms and molecules," Rev. Mod. Phys. \textbf{70} 1003 (1998), and references therein.

\bibitem{Jing07} H. Jing, J. Cheng and P. Meystre, "Coherent Atom-Trimer Conversion in a Repulsive Bose-Einstein Condensate," Phys. Rev. Letts. \textbf{99} 133002(2007).

\bibitem{ling05}H. Y. Ling, P. Maenner and H. Pu, "Coherent population trapping and dynamical instability in coupled atom-molecule condensates,"
 Phys. Rev. A, \textbf{72} 013608(2005).

\bibitem{ling07}H. Ling, P. Maenner, W. Zhang and H. Pu, "Adiabatic theorem for a condensate system in an atom-molecule dark state,"
Phys. Rev. A, \textbf{75} 033615(2007).

\bibitem{pu07}H. Pu, P. Maenner, W. Zhang and H. Ling, "Adiabatic Condition for Nonlinear Systems,"
Phys. Rev. Letts. \textbf{98} 050406 (2007).

\bibitem{kerman04} A. J. Kerman, J. M. Sage, S. Sainis, T. Bergerman and D. DeMille, "Production of Ultracold, Polar RbCs Molecules via Photoassociation," Phys. Rev. Letts. \textbf{92} 033004 (2004).

\bibitem{wang04} D. Wang, J. Qi, M. F. Stone, O. Nikolayeva, H. Wang, B. Hattaway, S. D. Gensemer, P. L. Gould, E. E. Eyler and W. C. Stwalley, "Photoassociative Production and Trapping of Ultracold KRb Molecules," Phys. Rev. Letts. \textbf{93} 243005 (2004).

\bibitem{sage05} J. M. Sage, S. Sainis, T. Bergeman and D. DeMille, "Optical Production of Ultracold Polar Molecules
," Phys. Rev. Letts. \textbf{94} 203001 (2005).

\bibitem{salzmann08} W. Salzmann, T. Mullins, J. Eng, M. Albert, R. Wester, and M. Weidem\"{u}ller, A. Merli, S. M. Weber, F. Sauer, M. Plewicki, F. Weise, "Coherent Transients in the Femtosecond Photoassociation of Ultracold Molecules
," L. W\"{o}ste, and A. Lindinger, Phys. Rev. Letts. \textbf{100} 233003 (2008).

\bibitem{viteau08} M. Viteau, A. Chotia, M. Allegrini, N. Bouloufa, O. Dulieu, D. Comparat and P. Pillet, "Optical Pumping and Vibrational Cooling of Molecules,"Science \textbf{321} 232 (2008).

\bibitem{deiglmayr08}J. Deiglmayr, A. Grochola, M. Repp, K. M\"{o}rtlbauer, C. Gl\"{u}ck, J. Lange, O. Dulieu, R. Wester and M. Weidem\"{u}ller, "Formation of Ultracold Polar Molecules in the Rovibrational Ground State
," Phys. Rev. Letts. \textbf{101} 133004 (2008).


\bibitem{winkler07} K. Winkler, F. Lang, G. Thalhammer, P. v. d. Straten, R. Grimm, and J. Hecker Denschlag, "Coherent Optical Transfer of Feshbach Molecules to a Lower Vibrational State
," Phys. Rev. Letts. \textbf{98} 043201 (2007).

\bibitem{lang08} F. Lang, K. Winkler, C. Strauss, R. Grimm, and J. Hecker Denschlag, "Ultracold Triplet Molecules in the Rovibrational Ground State
," Phys. Rev. Letts. \textbf{101} 133005 (2008).

\bibitem{danzl08} J. G. Danzl, E. Haller, M. Gustavsson, M. J. Mark, R. Hart, N. Bouloufa, O. Dulieu, H. Ritsch and H-C N\"{a}gerl, "Quantum Gas of Deeply Bound Ground State Molecules," Science \textbf{321} 1062 (2008).

\bibitem{ni08}K. K. Ni, S. Ospelkaus, M. H. G. deMiranda, A. Pe'er, B. Neyenhuis, J. J. Zirbel, S. Kotochigova, P. S. Julienne, D. S. Jin and J. Ye, "A High Phase-Space-Density Gas of Polar Molecules,"  Science \textbf{322} 231(2008).

\bibitem{ospelkaus08} S. Ospelkaus, A. Pe'er, K. K. Ni, J. J. Zirbel, B. Neyenhuis, S. Kotochigova, P. S. Jelienne, J. Ye and D. S. Jin, "Efficient state transfer in an ultracold dense gas of heteronuclear molecules," Nat. Phys. \textbf{4} 622 (2008).

\bibitem{takekoshi14}T. Takekoshi, L. Reichs\"{o}llner, A. Schindewolf, J. M. Hutson, C. Sueur, O. Dulieu, F. Ferlaino, R. Grimm and H-C N\"{a}gerl, "Ultracold Dense Samples of Dipolar Rb Cs Molecules in the Rovibrational and Hyperfine Ground State," Phys. Rev. Letts. \textbf{113} 205301(2014).

\bibitem{molony14} P. K. Molony, P. D. Gregory, Z. Ji, B. lu, M. P. K\"{o}ppinger, C. R. L. Sueur, C. L. Blackley, J. M. Hutson and S. L. Cornish, "Creation of Ultracold RbCs Molecules in the Rovibrational Ground State," Phys. Rev. Letts. \textbf{113} 255301 (2014).

\bibitem{park15} Jee Woo Park, Sebastian A. Will, and Martin W. Zwierlein, "Ultracold Dipolar Gas of Fermionic NaK Molecules in Their Absolute Ground State," Phys. Rev. Letts. \textbf{114} 205302 (2015).

\bibitem{kohler06} T. K\"{o}hler, K. G\'{o}ral and P. S. Julienne, "Production of cold molecules via magnetically tunable Feshbach resonances," Rev. Mod. Phys. \textbf{78} 1311 (2006).

\bibitem{chin10} C. Chin, R. Grimm, P. Julienne and E. Tiesinga, "Feshbach resonances in ultracold gases," Rev. Mod. Phys. \textbf{82} 1225 (2010).

\bibitem{zuchowski10} P. S. Zuchowski, J. Aldegunde and J. M. Hutson, "Ultracold RbSr Molecules Can Be Formed by Magnetoassociation," Phys. Rev. Letts. \textbf{105} 153201 (2010).

\bibitem{patel14} H. J. Patel, C. L. Blackley, S. L. Cornish and J. M. Hutson, "Feshbach resonances, molecular bound states, and prospects of ultracold-molecule formation in mixtures of ultracold K and Cs," Phys. Rev. A, \textbf{90} 032716 (2014).

\bibitem{wang15}F. Wang, X. He, X. Li, B. Zhu, J. Chen and D. Wang, "Formation of ultracold NaRb Feshbach molecules," New J. Phys. \textbf{17} 035003 (2015).

\bibitem{aikawa10} K. Aikawa,1 D. Akamatsu, M. Hayashi, K. Oasa, J. Kobayashi, P. Naidon, T. Kishimoto,
M. Ueda, and S. Inouye, "Coherent Transfer of Photoassociated Molecules into the Rovibrational Ground State
," Phys. Rev. Letts. \textbf{105} 203001 (2010).

\bibitem{stellmer12} Simon Stellmer,1 Benjamin Pasquiou, Rudolf Grimm, and Florian Schreck, "Creation of Ultracold Sr2 Molecules in the Electronic Ground State," Phys. Rev. Letts. \textbf{109} 115302 (2012).

\bibitem{shore91} B. W. Shore, K. Bergmann, J. Oreg and S. Rosenwaks, "Multilevel adiabatic population transfer," Phys. Rev. A, \textbf{44} 7442 (1991).


\bibitem{shore95} B. W. Shore, J. Martin, M. P. Fewell and K. Bergmann, "Coherent population transfer in multilevel systems with magnetic sublevels," Phys. Rev. A, \textbf{52} 566 (1995); Phys. Rev. A, \textbf{52} 583 (1995).

\bibitem{malinovsky97} V. S. Malinovsky and D. J. Tannor, "Simple and robust extension of the stimulated Raman adiabatic passage technique to N-level systems," Phys. Rev. A, \textbf{56} 4929 (1997).

\bibitem{vitanov98a} N. V. Vitanov, "Adiabatic population transfer by delayed laser pulses in multistate systems
," Phys. Rev. A \textbf{58} 2295 (1998).

\bibitem{vitanov98b} N. V. Vitanov, B. W. Shore and K. Bergmann, "Adiabatic population transfer in multistate chains via dressed intermediate states," Eur. Phys. J. D, \textbf{4} 15 (1998). 

\bibitem{vitanov99} N. V. Vitanov and S. Stenholm, "Adiabatic population transfer via multiple intermediate states
," Phys. Rev. A, \textbf{60} 3820 (1999).

\bibitem{sola99} Ignacio R. Sol\'{a}, Vladimir S. Malinovsky, and David J. Tannor, "Optimal pulse sequences for population transfer in multilevel systems
," Phys. Rev. A, \textbf{60} 3081 (1999).


\bibitem{jin04} S. Jin, S. Gong, R. Li and Z. Xu, "Coherent population transfer and superposition of atomic states via stimulated Raman adiabatic passage using an excited-doublet four-level atom," Phys. Rev. A, \textbf{69} 023408 (2004).

\bibitem{moller07} D. M\o ller, J. L. Sorensen, J. B. Thomsen and M. Drewsen, "Efficient qubit detection using alkaline-earth-metal ions and a double stimulated Raman adiabatic process," Phys. Rev. A, \textbf{76} 062321 (2007).

\bibitem{shapiro07} E. A. Shapiro, M. Shapiro, A. PeAer and J. Ye, "Photoassociation adiabatic passage of ultracold Rb atoms to form ultracold Rb2 molecules," Phys. Rev. A, \textbf{75} 013405 (2007).

\bibitem{qianNJP} J. Qian, L. Zhou, K. Zhang and W. Zhang, "Efficient production of polar molecular Bose--Einstein condensates via an all-optical R-type atom--molecule adiabatic passage," New J. Phys. \textbf{12} 033002 (2010).

\bibitem{qianJPB} J. Qian, K. Zhang, L. Zhou and W. Zhang, "Elimination of collisional effects in an R-type atom--molecule adiabatic passage," J. Phys. B, \textbf{43} 155206 (2010).

\bibitem{niu11} Ying-Yu Niu, Rong Wang, and Ming-Hui Qiu, "Stimulated Raman adiabatic passage in an extended ladder system," Phys. Rev. A, \textbf{84} 023406 (2011).

\bibitem{wang12} H. Wang, Y. M. Liu, J. W. Gao, D. Yan, R. Wang and J. H. Wu, "Adiabatic Raman passage via two-photon resonant transitions in a five-level system," Opt. Communs. \textbf{285} 3498 (2012).


\bibitem{kuznetsova08}E. Kuznetsova, P. Pellegrini, R. C\^{o}t\'{e}, M. D. Lukin and S. F. Yelin, "Formation of deeply bound molecules via chainwise adiabatic passage," Phys. Rev. A, \textbf{78} 021402R(2008).

\bibitem{qian10} J. Qian, W. Zhang and H. Ling, "Achieving ground-state polar molecular condensates by chainwise atom-molecule adiabatic passage
," Phys. Rev. A, \textbf{81} 013632 (2010).

\bibitem{danzl10}J. G. Danzl, M. J. Mark, E. Haller, M. Gustavsson, R. Hart, J. Aldegunde, J. M. Hutson and H-C N\"{a}gerl, "An ultracold high-density sample of rovibronic ground-state molecules in an optical lattice," Nat. Phys. \textbf{6} 265(2010).

\bibitem{feng09} Z. Feng, W. Li, L. Wang, L. Xiao and S. Jia, "Optimum conditions for producing Cs2 molecular condensates by stimulated Raman adiabatic passage," Phys. Rev. A, \textbf{80} 043620 (2009).

\bibitem{javanainen99}J. Javanainen  and M. Mackie, "Coherent photoassociation of a Bose-Einstein condensate," Phys. Rev. A, \textbf{59} R3186 (1999).

\bibitem{mackie00}M. Mackie, R. Kowalski and J. Javanainen, "Bose-Stimulated Raman Adiabatic Passage in Photoassociation," Phys. Rev. Letts. \textbf{84} 3803 (2000).

\bibitem{wynar02}P. D. Drummond, K. V. Kheruntsyan, D. J. Heinzen and R. H. Wynar, "Stimulated Raman adiabatic passage from an atomic to a molecular Bose-Einstein condensate," Phys. Rev. A, \textbf{65} 063619(2002).

\bibitem{naidon03} P. Naidon and F. M-Seeuws, "Pair dynamics in the formation of molecules in a Bose-Einstein condensate," Phys. Rev. A, \textbf{68} 033612 (2003).

\bibitem{kallush08} S. Kallush and R. Kosloff, "Unitary photoassociation: One-step production of ground-state bound molecules," Phys. Rev. A, \textbf{77} 023421 (2008).

\bibitem{heinzen00}D. J. Heinzen, R. Wynar, P. D. Drummond and K. V. Kheruntsyan, "Superchemistry: Dynamics of Coupled Atomic and Molecular Bose-Einstein Condensates," Phys. Rev. Letts. \textbf{84} 5029 (2000).

\bibitem{oreg92}J. Oreg, K. Bergmann, B. W. Shore and S. Rosenwaks, "Population transfer with delayed pulses in four-state systems," Phys. Rev. A, \textbf{45} 4888 (1992).

\bibitem{meng14} S. Y. Meng, X. H. Chen, S. N. Ning, J. M. Wen and L. B. Fu, "Instability, adiabaticity and controlling effects of external fields for the dark state in a heteronuclear atom--tetramer conversion system ," J. Phys. B, \textbf{47} 185303 (2014).

\bibitem{jia12} N. Jia, J. Qian, G. Dong and W. Zhang, "Stability, adiabaticity and transfer efficiency in a nonlinear $\Lambda$-system," J. Phys. B, \textbf{45} 015301(2012).

\bibitem{wynar00} R. Wynar, R. S. Freeland, D. J. Han, C. Ryu and D. J. Heinzen, "Molecules in a Bose-Einstein Condesate", Science, \textbf{287} 1016 (2000).

\bibitem{kirilov15} E. Kirilov, M. J. Mark, M. Segl and H. C. N\"{a}gerl, "Compact, robust, and spectrally pure diode-laser system with a filtered output and a tunable copy for absolute referencing", Appl. Phys. B \textbf{119} 233 (2015).

\bibitem{yatsenko14} L. P. Yatsenko, B. W. Shore and K. Bermann, "Detrimental consequences of small rapid laser fluctuations on stimulated Raman adiabatic passage", Phys. Rev. A., \textbf{89} 013831 (2014).


\end{thebibliography}
\end{document}